\documentclass{article}
\usepackage{amssymb}
\usepackage{amsfonts}

\newtheorem{theorem}{Theorem}

\newtheorem{definition}[theorem]{Definition}

\input{tcilatex}
\begin{document}

\title{On the emergence of duration from quantum observation and some
consequences}
\author{Andreas Schlatter, Burghaldeweg 2F, 5024 K\"{u}ttigen Switzerland}
\maketitle

\begin{abstract}
We show that by the act of quantum measurement on a system there emerges a
notion of duration and a corresponding time flow which turns out to be the
thermal flow connected to the system. We show that, under some
quasi-classical assumptions on the observer, this flow shows relevant
properties of empirical time and some interesting consequences for Special
Relativity are drawn.

\end{abstract}

\bigskip

\section{Introduction}

In both theories, Special Relativity (SR) and Quantum Mechanics (QM), the
observer plays an important role. In SR this has, amogst others, the known
implication that the question "which clock goes faster?" is not well posed,
if one considers the clocks of two observers in relative uniform motion. In
SR time and space merge to a continuum whose isometry group (Lorentz
transformations) mixes space and time coordinates. SR finds it as difficult
as Newtonian physics does, to explain some empirical features of time like,
for example, its direction and duration. It can be shown in classical
physics, including General Relativity, that coordinate time can be made
disappear altogether as a fundamental quantity and what is left is change of
physical quantities in relation to other physical quantities $\left[ 1,8%
\right] .$Traditional QM is even more radical in establishing a dependency
on the observer. An observer must interact with a system to form
correlations which then determine probabilities for different outcomes of a
specific observable.This view is systematically developed in $\left[ 2\right]
.$In the corresponding relational interpretation of QM correlations between
systems and observers through interaction are considered a complete
description of the world. In relation to time, however, QM makes in a way a
step back and uses the absolute time concept of Newtonian physics.

In the present paper we want to make use of the relational approach to QM
and develop a notion of time, and in particular duration, which emerges from
(quantum) observation.The corresponding time flow will be shown to have some
properties of empirical time and to allow interesting connections to the
notion of time in SR.

\section{Duration}

QM is about establishing probabilities for measurement-outcomes of
observable physical properties. It is an empirical fact that repeated
measurements result in a chain of single valued outcomes, i.e.
manifestations, of these properties.Time is not part of the quantum
mechanical framework but enters the theory as an external parameter.It is
known not to be an observable represented by an operator under realistic
physical assumptions, i.e. a bound on energy from below.To include time, we
take the route to ask the question, how long exactly does it take to
complete a measurement. An answer to the question about the duration of a
measurement has been given in the spirit of QM in $\left[ 3\right] .$This
will be our starting point.

For the sake of simplicity, we assume in the sequel to be in a two
dimensional space-time with one real "time" coordinate $t\in 
\mathbb{R}
$ and one real space coordinate $x\in 
\mathbb{R}
$. We also make the assumption of a (global) temperature $T$. Let a quantum
observer \emph{O}\textit{\ }have knowledge, through prior measurement, of
the initial state $\psi _{0}$ of a system \emph{S}\textit{\ }and of an
initial state $A_{0}$ of an apparatus \emph{A}.The observer \emph{O}\textit{%
\ }must interact with \emph{A}\textit{\ }to read off the information which
the apparatus has about \emph{S}. By doing that, \emph{O}\textit{\ }makes a
measurement and will observe only one branch of the evolution of the \emph{AS%
}\textit{-}system to get a definite result.Without interaction, \emph{O}%
\textit{\ }will describe a unitary evolution $\Xi $ of correlations between 
\emph{S}\textit{\ }and apparatus \emph{A}, $\Xi :=\psi _{0}\rangle \otimes
A_{0}\rangle \rightarrow \psi _{I}\rangle \otimes A_{I}\rangle ,0\rightarrow 
\overline{t},$ where $A_{I}\rangle $ and $\psi _{I}\rangle $ are a
respective orthogonal basis of apparatus and system, relating to some fixed
observable, and $I$\textit{\ }denotes the set of indices over the spectrum.
By reduction over \emph{A}\textit{, \emph{O} }assigns to \emph{S}\textit{\ }%
at $\overline{t}$ the density matrix $\Psi .$ The corresponding entropy is $%
S=-k\cdot tr\left( \Psi \log \Psi \right) .$The question "within what time
has the measurement happened?" can be answered in an operative fashion $%
\left[ 3\right] .$ Define the operator

\[
M:=\dsum\limits_{I}\left( \psi _{I}\rangle \otimes A_{I}\rangle \right)
\left( \psi _{I}\rangle \otimes A_{I}\rangle \right) . 
\]%
There holds $M\left( \psi _{I}\rangle \otimes A_{J}\rangle \right) =\delta
_{IJ}\cdot Id.$ $M$\textit{\ }is a self-adjoint operator on the Hilbert
space of the coupled system \emph{AS}\textit{\ }and therefore represents an
observable. $M=1$ does imply that the pointer indicates the correct value of 
\emph{S}, where $M=0$ means it does not. We now say that, if the pointer
correctly indicates the value, the measurement has happened, else it has
not. In particular, if we follow the Schr\"{o}dinger evolution $\Xi $ of the
coupled system from $0$ to $\overline{t}$, we can define the probability $P$%
\textit{\ }that the measurement has happened by $P\left( t\right) =\langle
\Xi \left( t\right) \left\vert M\right\vert \Xi \left( t\right) \rangle .$
Correspondingly, there is entropy $S=-k\cdot tr\left( \Psi \log \Psi \right) 
$\textit{\ }with probability $P\left( t\right) $\textit{.}The question
"within what time has the measurement happened?" is therefore not well posed
in the classical sense but finds an answer in the probabilistic sense of QM.
We realize that this "duration" is a relational quantity between two
different states of the system-apparatus complex \emph{AS}. It is invariant
under unitary transformations of \emph{AS} and is hence a full observable,
namely measurable and predictable.

The stream of measurement outcomes is realized in the spectrum of the
observable and a one-dimensional parameter, called (coordinate) time, which
just represents the basic (empirical) fact that there is single valued
change of spectral values which can be labelled by this parameter. The main
challenge of the above answer to the \textit{"within what time"}-question
is, that duration is not well defined. For an observer \emph{O}\textit{\ }a
"second" resulting from one measurement needs not be a "second" resulting
from the next one. Another challenge is that the time interval not only
depends upon the system \emph{S }but also on the apparatus \emph{A} and the
interaction between the two. To overcome these difficulties we have to
define a specific observer.

\section{The quasi-classical observer}

\begin{definition}
Let for some measurement evolution $\Theta \left( t\right) $, $P\left(
t\right) =\langle \Theta \left( t\right) \left\vert M\right\vert \Theta
\left( t\right) \rangle .$ With respect to the measurement of a specific
observable on a quantum system \emph{S}\textit{\ }with evolution $\Theta
\left( t\right) $, observer \emph{O} is quasi-classical, if \textit{i) }$%
P\in \left\{ 0,1\right\} ,$\textit{ii) OS is a separable state.\newline
}
\end{definition}

We will now construct a specific quasi-classical observer. Let us look at
the entropy balance sheet of the total system ASO. As mentioned above, the
correlation between A\ and S\ leads, at $\overline{t}$, with certainty to
entropy of the amount $S=-k\cdot tr\left( \Psi \log \Psi \right) $, where $k$
is the Boltzmann factor and $\Psi $ the mixed state which O\ assigns to S\
by reduction over A. For any intermediate $0\leq t\leq \overline{t}$ it
exists with a certain probability $P\left( t\right) $. This entropy must
also be there from the perspective of the inner observer A\ (apparatus),
after having registered a definite result, in order to preserve the second
law of thermodynamics $\left[ 4\right] .$ For this reason it was possible
after all to discover the second law in the 19th century, because it relates
to the "stream of consciousness" of an inner observer who experiences the
world being in definite states. From the perspective of O, there is entropy
created because of the correlations between A\ and S. From the perspective
of A there is entropy creation because the former state of the register has
been erased (forgotten) by acquiring a new definite result. Given the
environment $\Omega $ at temperature $T,$ this erasure induces an average
dissipation of energy $\overline{E}=-kTtr\left( \Psi \log \Psi \right) $
into the environment $\Omega .$ By construction, O does not know the full
dynamics $H$\ related to the dissipation. What O does know, however, it the
corresponding average energy $\overline{E}$ which is sufficient for our
purposes as we will see below. The Hamiltonian $H$\ acts on $\Omega \otimes
AS$\ leading to the sequence%
\begin{equation}
\omega _{0}\rangle \otimes AS_{0}\rangle \rightarrow \omega _{I}\rangle
\otimes AS_{I}\rangle ,\;\left\langle \omega _{0},\omega _{I}\right\rangle
=0.  \label{1}
\end{equation}%
We now define $O\subset \Omega $\textit{\ }to be quasi-classical with
respect to the dissipation-induced evolution (1). By Def 1. i), it follows
that \emph{O}\textit{\ }has perceived the measurement, exactly if the state $%
\omega _{I}\rangle $ is reached.The time $\Delta t$ it takes to do this we
chose to be the minimal coordinate-time interval $\Delta t_{\min }$ it takes 
$\Omega $ to evolve from its initial state $\omega _{0}\rangle $ indicating
"nothing has happened" into an orthogonal state $\omega _{I}\rangle $
indicating "measurement completed with certainty". So we actually look at
the (minimal) time that elapses until \emph{O}\textit{\ }has perceived that
there is a definite correlation between apparatus and system.The necessary
interaction between \emph{AS}\textit{\ }and the environment is triggered by
the dissipation of entropy/energy once \emph{A}\textit{\ }has erased a
former state \emph{A}$_{0}.$

Is there a way to have an indication how long this takes? The answer is
"yes" because of a well-established fact by Margolus and Levitin,$\left[ 6%
\right] $, that a general quantum system with average energy $\overline{E}$
(assuming a ground state $E_{0}=0$) takes at least $\Delta t_{\min }=\frac{h%
}{4\overline{E}}$ to evolve into an orthogonal state ($h$\textit{\ }is
Planck's constant).Therefore we derive%
\begin{equation}
\Delta t_{\min }=\frac{h}{4\overline{E}}=\frac{h}{4kTS}.  \label{2}
\end{equation}%
The minimum is achievable if $2\overline{E}$ is an element of the energy
spectum of $H$\textit{\ }and will nearly be attained in macroscopic systems,
where we would expect spectral values to come close to $2\overline{E}$.
Because $\overline{E}=-kTtr\left( \Psi \log \Psi \right) $ we realize, that
the time interval which observer \emph{O}\textit{\ }experiences in
connection with the measurement on \emph{S}\textit{\ }can be derived from
the Schr\"{o}dinger flow generated by the Hamiltonian $H=-kT\log \Psi $
induced by the system.This flow and its time quantum $\Delta t_{\min }$ only
depend on the system \emph{S.\footnote{%
We could not simply have set $\Delta t_{\min }=\overline{t}$ and focused
only on the \emph{AS }interaction, since this would have implied a
dependence on the apparatus.}}The flow of time is hence a series of these
time quanta. There is no time elapsing, if there is nothing "happening" and
hence no further dissipation to process states.

Since \emph{O }is defined to be quasi-classical, the \emph{OS }system is
separable. Assume it has the general form of a classical-quantum state%
\begin{equation}
\Psi ^{OS}=\dsum\limits_{i\in I}p^{O}\left( i\right) i\rangle \langle
i\otimes \Psi _{i}^{S},  \label{3}
\end{equation}%
where $p^{O}\left( i\right) $ are the classical probabilities of some
distinguishable states $i\rangle $ of \emph{O. }This is a plausible
assumption, especially in the light of SR. We may now look at the time flow
as seen by a further observer \emph{P }who\emph{\ }knows \emph{S }and \emph{O%
}. In order to determine the time flow from this perspective, we use the
formula of conditional entropy for a classical-quantum state [7]

\begin{equation}
S\left( S\mid O\right) :=S\left( S,O\right) -S\left( O\right)
=\dsum\limits_{i\in I}p^{O}\left( i\right) S\left( \Psi _{i}^{S}\right) ,
\label{4}
\end{equation}%
where $S\left( S,O\right) $ denotes the joint entropy.There always holds $%
S\left( S\mid O\right) \leq S\left( S\right) $ . If the inequality is
strict, then there is a sort of time dilation between the time flows $\Delta
t_{\min }^{S\mid O}>\Delta t_{\min }^{S}$. As soon as the notion of
separability is abandoned, there are challenges to be expected. If the
correlation between \emph{O} and \emph{S} were purely quantum, then the
relative entropy might become negative \ $S\left( S\mid O\right) <0$. In
this case the expected energy of dissipation is negative $\overline{E}<0$ .
We will dedicate the last paragraph 5 to the case of these anti-qubits.

To summarize, we have constructed a quasi-classical observer who (passively)
notices what is happening to separated physical systems by registering
definite outcomes of experiments. This observer experiences a time flow
moving in discrete steps, defined by the system \emph{S} with correspondig
density matrix $\Psi $.This flow turns out to be the thermal time flow for $%
\Psi $,$\left[ 8\right] $, with \textquotedblleft quanta\textquotedblright\ $%
\Delta t_{\min }^{S}$,which get smaller, if the average energy per bit of
information $kT=\frac{\partial E}{\partial S}$ gets bigger. Since the
dissipation of information into the environment, which triggers the flow, is
irreversible, the time flow has a direction. The time quanta $\Delta t_{\min
}^{S}$ are time steps which remain the same over repeated measurements and
therefore allow constant ratios

\[
\frac{\Delta t_{\min }^{S_{1}}}{\Delta t_{\min }^{S_{2}}}\newline
\]%
between thermal flows with respect to two systems $S_{1},S_{2}$.The marching
in step of \textquotedblleft natural clocks\textquotedblright\ in that sense
is thus possible.We will now choose a quasi-classical position observer,
construct the corresponding thermal-flow and discuss some consequences.

\section{Consequences for space-time}

\subsection{Bound on velocity}

For a non-interacting observer with (coordinate) time parameter $t$ , the
system \emph{S}, with initial state $\psi _{0}$, undergoes a unitary
evolution $\psi \left( t\right) ,t\geqq 0$. Entropy is an invariant quantity
under the time evolution and there holds $S_{0}:=S\left( \psi _{0}\right)
=S\left( \psi \left( t\right) \right) .$It is now possible to introduce the
velocity of the process of repeated measurement over a period $\left[ 0,t%
\right] $. If we denote by $\theta =\frac{4kTS_{0}}{h}\cdot t$ the number of
orthogonal states which the system passes over an interval of (coordinate)
time $\left[ 0,t\right] $, we can define the \textquotedblleft
velocity\textquotedblright\ by

\begin{equation}
v=\frac{\theta }{t}=\frac{4kT}{h}\cdot S_{0}.  \label{5}
\end{equation}%
This corresponds to how fast the observer is registering consecutive
measurement outcomes of the system with initial field $\psi _{0}$. Please
note that for the time parameter in $\left( \ref{5}\right) $ we may use any
(coordinate) time parameter $s$.

We now want to apply $\left( \ref{5}\right) $ to the special situation where
position is measured. Remember, we assumed to be in a two dimensional
space-time with one real \textquotedblleft time\textquotedblright\
coordinate $t\in 
\mathbb{R}
$ and one real space coordinate $r\in 
\mathbb{R}
$.We choose $\psi _{0}$ to be a free particle, represented by a Gaussian
wave package, with mean wave number $\langle k\rangle =k_{0}$ and particle
source $\langle r\rangle $ at the origin

\qquad \qquad \qquad \qquad 
\[
\psi _{0}=\sqrt[2]{\frac{2\sigma _{0}}{\sqrt[2]{\pi }}}\exp \left( -\frac{%
\sigma _{k_{0}}^{2}r^{2}}{2}\right) \exp \left( ik_{0}r\right) . 
\]%
A quasi-classical observer, asking whether the particle is inside or outside
an interval $\left[ 0,R\right] $, decomposes $\psi _{0\text{ }}$into two
orthogonal fields

\[
\psi _{0}=\psi _{0}\cdot Id_{\left[ 0,R\right) }+\psi _{0}\cdot Id_{\left[
R,\infty \right) }=\psi _{R}+\psi _{R}^{\perp }. 
\]%
The two fields are eigenvectors to the corresponding projectors $Id_{R}$ and 
\newline
$\left( 1-Id_{R}\right) $.The corresponding density matrix is

\[
\Psi _{R}=\left( 
\begin{tabular}{ll}
$\left\vert \psi _{R}\right\vert ^{2}$ & $0$ \\ 
$0$ & $\left\vert \psi _{R}^{\perp }\right\vert ^{2}$%
\end{tabular}%
\right) . 
\]
For entropy we get

\[
S\left( \Psi _{R}\right) =-\left[ 
\begin{array}{c}
\left( \dint\limits_{0}^{R}\left\vert \psi _{R}\right\vert ^{2}dr\right)
\left( \log \left( \dint\limits_{0}^{R}\left\vert \psi _{R}\right\vert
^{2}dr\right) \right) + \\ 
+\left( \dint\limits_{R}^{\infty }\left\vert \psi _{R}^{\perp }\right\vert
^{2}dr\right) \left( \log \left( \dint\limits_{R}^{\infty }\left\vert \psi
_{R}^{\perp }\right\vert ^{2}dr\right) \right) 
\end{array}%
\right] .
\]%
Note that the probability to find a particle within $\left[ 0,R\right] $ is
increasing with $R$ , since $\dint\limits_{R}^{\infty }\left\vert \psi
_{R}^{\perp }\right\vert ^{2}dr\rightarrow 0,R\rightarrow \infty $.

We introduce the error function $\func{erf}\left( r\right) :=\frac{2}{\sqrt[2%
]{\pi }}\dint\limits_{0}^{r}e^{-t^{2}}dt$ to get

\begin{eqnarray*}
S\left( \Psi _{R}\right) &=&-\left[ 
\begin{array}{c}
\func{erf}\left( \frac{R}{\sigma _{k_{0}}}\right) \log \left( \func{erf}%
\left( \frac{R}{\sigma _{k_{0}}}\right) \right) + \\ 
+\left( 1-\func{erf}\left( \frac{R}{\sigma _{k_{0}}}\right) \right) \log
\left( 1-\func{erf}\left( \frac{R}{\sigma _{k_{0}}}\right) \right)%
\end{array}%
\right] \\
&:&=-G\left( \frac{R}{\sigma _{k_{0}}}\right) .
\end{eqnarray*}%
\newline
In the function $G$ the parameter $\sigma _{k_{0}}$ only appears in the
argument and the maximum $C$ over $\left[ 0,R\right] $ is therefore
independent of $\sigma _{k_{0}}$. It can be computed and turns out to be $%
C=\log 2$ , over $\left[ 0,\infty \right) .$ Therefore a repeated
measurement of the particle within any interval $\left[ 0,R\right] $ cannot
go faster in the sense of $\left( \ref{5}\right) $ than

\begin{equation}
v_{\max }\leq 4\log 2\cdot \frac{kT}{h}.\footnote{%
The particle could lie in $\left[ R,\infty \right) ,$ albeit with a small
probability for large $R.$}  \label{7}
\end{equation}%
Estimate $\left( \ref{7}\right) $ is independent of the line element $dr$. A
quasi-classical observer will thus \textquotedblleft see\textquotedblright\
the particle moving with at most a process velocity of $v_{\max }$ .

If we chose the measure to express the number of orthogonal states, i.e.
positions in an interval, to be the distance $\Delta x,$ we can go a step
further and consider not only process velocity but classical (average)
velocity as defined by $v:=\frac{\Delta x}{\Delta t}\leq \frac{R}{\Delta
t_{\min }}=\frac{4kT}{h}S\left( \Psi _{R}\right) \cdot R$ , if the particle
materializes within $\left[ 0,R\right] $. We again make use of the error
function to get

\begin{eqnarray*}
S\left( \Psi _{R}\right) \cdot R &=&-\left[ 
\begin{array}{c}
\func{erf}\left( \frac{R}{\sigma _{k_{0}}}\right) \log \left( \func{erf}%
\left( \frac{R}{\sigma _{k_{0}}}\right) \right) + \\ 
+\left( 1-\func{erf}\left( \frac{R}{\sigma _{k_{0}}}\right) \right) \log
\left( 1-\func{erf}\left( \frac{R}{\sigma _{k_{0}}}\right) \right)%
\end{array}%
\right] \cdot R \\
&:&=-G\left( \frac{R}{\sigma _{k_{0}}}\right) \cdot R=-G\left( \frac{R}{%
\sigma _{k_{0}}}\right) \cdot \frac{R}{\sigma _{k_{0}}}\cdot \sigma
_{k_{0}}:=H\left( \frac{R}{\sigma _{k_{0}}}\right) \cdot \sigma _{k_{0}}.
\end{eqnarray*}%
In the function $H$ the parameter $\sigma _{k_{0}}$ also only appears in the
argument and the maximum of the function over $\left[ 0,R\right] $ is
therefore independent of $\sigma _{k_{0}}$. It can be computed and turns out
to be $0.4579.$ By the definition we therefore derive

\begin{equation}
v_{\max }\leq 1.832\cdot \frac{kT}{h}\cdot \sigma _{k_{0}}.  \label{8}
\end{equation}%
With the uncertainty relation $\sigma _{x_{0}}\cdot \sigma _{k_{0}}=1$ , we
have\qquad \qquad \qquad \qquad 
\begin{equation}
v_{\max }\leq \frac{kT}{h\sigma _{x_{0}}}.  \label{9}
\end{equation}%
For free (Gaussian) particles there is therefore a bound on classical
velocity independent of the initial average momentum $k_{0}$. In fact we see
that estimate $\left( \ref{9}\right) $ implies that there is a universal
bound on velocity if and only if there is a lower bound on the resolution of
space $\sigma _{x_{0}}\geq \Lambda _{0}$ and that

\begin{equation}
v_{\max }\leq \frac{kT}{h\Lambda _{0}}.  \label{a}
\end{equation}

\subsection{Consistency}

Classically, space-time coordinates serve the dual purpose to label
individual objects and to represent physical properties of these. A change
in observer equals a change in gauge (co-ordinates).The experimentally
observed invariance of a specific speed must then be reflected by invariance
under the transformation group of the co-ordinate system with regard to
moving frames. The corresponding transformation group is the Lorentz group $%
\Lambda .\footnote{%
The existence of a frame-independent velocity $v_{\max }$ is only one
postulate of SR, the other one being the general covariance of physical laws
with repect to change of Lorentz frame.}$Consistency would leave us to
expect that $v_{\max }$ $\left( \ref{8}\right) $ , which we have logically
derived, not empirically found and which is a composite quantity, is
invariant under Lorentz transformations.

Remember the set up: a quasi-classical observer \emph{O} measures position
within an interval $\left[ 0,R\right] $ of a system \emph{S} with field $%
\psi _{0}$ in form of a Gaussian wave packet.The corresponding density
matrix is $\Psi _{R}$ . Assume there is a frame with another quasi-classical
observer $\overline{O}$ that moves with velocity $v$ in the negative $x$%
-direction. From the perspective of \ $\overline{O}$ the \emph{OS} system
moves in the positive $x$-direction and is, for any $R$, a classical-quantum
system of the form $\Phi _{OS}=v\rangle \langle v\otimes \overline{\Psi }%
_{R}.$ By $\left( \ref{4}\right) $ we derive that

\begin{equation}
S\left( S\mid O\right) =S\left( \overline{\Psi }_{R}\right) .  \label{10}
\end{equation}%
A field function $\overline{\psi }=\overline{\psi }\left( \overline{x}%
\right) $ must reflect that motion and the $\overline{O}$ co-ordinates, $%
\overline{x}$, are correlated with the moving ones, $x$ , by
Lorentz-transformations $x=\gamma \overline{x}+a$, where $\gamma =\frac{1}{%
\sqrt[2]{\left( 1-\frac{v^{2}}{c^{2}}\right) }}$and $a\in 
\mathbb{R}
$.The length of any object,$r$, is contracted and there holds the
transformation $\overline{r}=\gamma ^{-\frac{1}{2}}r$. We want to fall back
to the original way of looking at relativistic thermodynamics in $\left[ 9%
\right] $, where temperature transforms according to

\begin{equation}
T=\gamma ^{-\frac{1}{2}}\overline{T}.  \label{11}
\end{equation}%
For entropy $S$ we calculate the transformation law by plugging the
transformations into the definition of $S$ to get

\begin{equation}
S:=S\left( r\right) =\overline{S}\left( \gamma \overline{r}\right) :=%
\overline{S}.  \label{12}
\end{equation}%
With $\left( \ref{4}\right) $ there holds

\begin{equation}
\Delta t_{\min }=\frac{h}{4kTS}=\frac{\gamma h}{4k\overline{T}\overline{S}}%
=\gamma \overline{\Delta t}_{\min }.  \label{13}
\end{equation}%
Note that we assume that the fundamental constants of nature $h,k$ are
invariant. By the length-contraction $\overline{r}=\gamma ^{-1}r$ , observer 
$\overline{O}$ thus derives

\begin{equation}
\frac{\overline{r}}{\overline{\Delta t}_{\min }}=\frac{\gamma ^{-1}r}{\gamma
^{-1}\Delta t_{\min }}\leq 1.832\frac{kT}{h}\cdot \sigma _{k_{0}}.
\label{14}
\end{equation}%
The bound is indeed invariant.

\subsection{Simultaneity}

So far we addressed the question of duration which turned out to be a full
observable. We found that there is indeed an element of minimal duration or
time quantum related to a thermal time flow in case an observer is
quasi-classical. Another question is the one of simultaneity of two events.
If a quasi-classical observer \emph{O }measures two systems and two
different observables it is very diffcult to give an operative definition of
simultaneity, since we compare pairs with apples. If the measurements relate
to one single observable and there is \ maximal process velocity $v_{\max },$%
then there is the possibility of the following operational definition. An
event corresponds to the passing of one state into a different, orthogonal
one. Given two time intervals $\left[ 0,t_{1}\right] $, $\left[ 0,t_{2}%
\right] ,$ the number of orthogonal states which are processed by repeated
events within the intervals are $\theta _{1}=\frac{4kTS_{1}}{h}\cdot t_{1}$
and $\theta _{2}=\frac{4kTS_{2}}{h}\cdot t_{2}$ , respectively. Assuming
that the process velocity was maximal, we can define the two starting points
of the two processes to be simultaneous, if the time difference $\Delta t$
of their measured end by \emph{O }is

\begin{equation}
\Delta t=\frac{\theta _{2}-\theta _{1}}{v_{\max }}.  \label{c}
\end{equation}%
\newline
In case of position measurement and classical velocity, $\left( \ref{c}%
\right) $ just says that two distant events are simultaneous with repect to 
\emph{O, }if the difference of arrival times of two light beams sent at the
moment of the events amount to the difference of their respective distance
from \emph{O }divided by the speed of light.

\section{Anti-qubits}

In paragraph 2 we shortly mentioned the case of anti-qubits which we will
now follow up in some detail. In this paragraph we denote the density matrix 
$\Psi $ of a state by $\rho _{\Psi }$. In the case of a single state $\psi $
the density matrix $\rho _{\Psi }=\psi \rangle \langle \psi $ is positive
semi-definite, $\rho _{\Psi }\geq 0$ , and entropy is well defined. In case
of a composite system consisting of two states $\Psi $ and $\Phi $ and with
corresponding joint density matrix $\rho _{\Psi \Phi }$ we can define the
generalized conditional entropy\newline
\[
S\left( \Psi \mid \Phi \right) :=-k\cdot tr\left( \rho _{\Psi \Phi }\log
\rho _{\Psi \mid \Phi }\right) ,\newline
\]
where\newline
\[
\rho _{\Psi \mid \Phi }=\lim_{n\rightarrow \infty }\left[ \rho _{\Psi \Phi
}^{\frac{1}{n}}\left( id_{\Psi }\otimes \rho _{\Phi }\right) ^{-\frac{1}{n}}%
\right] ^{n}\newline
. 
\]%
It turns out,$\left[ 5\right] ,$that conditional entropy is well defined for
any composite system and that $S\left( \Psi \mid \Phi \right) =S\left( \Psi
,\Phi \right) -S\left( \Phi \right) $. If the two states are separable, 
\newline
\[
\rho _{\Psi \Phi }=\dsum\limits_{i\in I}w_{i}\rho _{\Psi }^{i}\otimes \rho
_{\Phi }^{i},0\leq w_{i}\leq 1,\dsum\limits_{i\in I}w_{i}=1,\newline
\]
then there holds that the conditional entropy is positive $S\left( \Psi \mid
\Phi \right) \geq 0$ . This ensures that the concept of \textquotedblleft
thermal\textquotedblright\ time works for a quasi-classical observer.In case
of entangled systems, however, there might hold $S\left( \Psi \mid \Phi
\right) <0.$ The average energy dissipated would then be negative and, in
analogy to the situation with anti-particles, the question arises whether
such anti-qubits $\rho _{\Psi \mid \Phi }$ move backwards in the
thermal-time defined by the Hamiltonian $H=-kT\log \rho _{\Psi \mid \Phi }$.

It turns out that the Margolus-Levitin theorem is \textquotedblleft
robust\textquotedblright\ with respect to density matrices with spectrum
bounded from below. Assume that the eigenvalues $E_{i},i\in I$ of $H$ are
bounded from below, i.e. $E_{i}\geq E_{0}$, which is the only physically
realistic situation. There holds

\begin{equation}
\Delta t_{\min }^{\rho _{\Psi \mid \Phi }}\geq \frac{h}{4\left( \overline{E}%
-E_{0}\right) }\geq 0.  \label{15}
\end{equation}%
Inequality $\left( \ref{15}\right) $ results directly from the proof of the
Margolus-Levitin theorem $\left[ 3\right] $ . Assume that $\rho _{\Psi \mid
\Phi }=\dsum\limits_{n}c_{n}\mid E_{n}\rangle $ evolves under $H$ to \newline
\[
\rho _{t}=\dsum\limits_{n}c_{n}e^{-\frac{i}{\hbar }E_{n}t}\mid E_{n}\rangle
. 
\]%
\newline
There holds%
\[
\langle \rho _{\Psi \mid \Phi }\mid \rho _{t}\rangle
=\dsum\limits_{n}\left\vert c_{n}\right\vert ^{2}e^{-\frac{i}{\hbar }%
E_{n}t}=e^{\frac{i}{\hbar }E_{0}t}\dsum\limits_{n}\left\vert
c_{n}\right\vert ^{2}e^{-\frac{i}{\hbar }\left( E_{n}-E_{0}\right) t}. 
\]%
To find the smallest value of $t$ such that $\langle \rho _{\Psi \mid \Phi
}\mid \rho _{t}\rangle =0$, we write

\begin{eqnarray}
\func{Re}\left( \langle \rho _{\Psi \mid \Phi }\mid \rho _{t}\rangle \right)
&=&\dsum\limits_{n}\left\vert c_{n}\right\vert ^{2}\cos \left( \frac{\left(
E_{n}-E_{0}\right) }{\hbar }\right) \geq  \label{b} \\
&\geq &\dsum\limits_{n}\left\vert c_{n}\right\vert ^{2}\left( 1-\frac{2t}{%
\pi }\left( \frac{E_{n}-E_{0}}{\hbar }\right) \right) +\sin \left( \frac{%
\left( E_{n}-E_{0}\right) }{\hbar }t\right) =  \nonumber \\
&=&1-\frac{2\left( \overline{E}-E_{0}\right) }{\pi \hbar }t+\frac{2}{\pi }%
\func{Im}\left( \langle \rho _{\Psi \mid \Phi }\mid \rho _{t}\rangle \right)
.\newline
\nonumber
\end{eqnarray}%
$\newline
$Inequality $\left( \ref{b}\right) $ follows because of $\cos \left(
x\right) \geq 1-\frac{2}{\pi }\left( x+\sin x\right) ,x\geq 0.$\newline
If $\langle \rho _{\Psi \mid \Phi }\mid \rho _{t}\rangle =0,$ then both $%
\func{Re}=0$ and $\func{Im}=0$, and there follows $\left( \ref{15}\right) $.
Therefore, anti-qubits also move forwards in thermal-time and for the
process velocity $\left( \ref{5}\right) $ there holds

\begin{equation}
\frac{\theta }{t}=\frac{4kT\left( S\left( \Psi \mid \Phi \right) -\left(
\log \rho _{\Psi \mid \Phi }\right) _{0}\right) }{h}.  \label{17}
\end{equation}

Here $\left( \log \rho _{\Psi \mid \Phi }\right) _{0}$ is the first (lowest)
eigenvalue of $H=-\log \rho _{\Psi \mid \Phi }.$

\section{Conclusion}

In the thermal time flow which we derived by defining a quasi-classical
observer, there exists a quantum of duration which gives the
\textquotedblleft now\textquotedblright\ an extension. The
"quasi-classicality" consists of the idea that the end-observer is only able
to perceive an indirect confirmation that the measurement has happened in
form of distinguishable states of the environment which he is part of. The
transition between these orthogonal states happens by a
Schroedinger-evolution governed by entropy dissipation which is triggered by
the measurement. The resulting time flow only depends on the original system
which is an important feature of the classical description.

What are the prerequisites to derive the (thermal) flow? The flow results
from the perception of change of some physical property of a system. If no
(external) observer can be defined, like in the case where the system is the
whole universe, there can be no single time flow in the above sense to
describe the change of the whole system. The question whether this is a
limiting condition is another one, since it is known that the state of a
system is uniquely defined by the correlations of its subsystems $\left[ 10%
\right] $.The flow further depends upon the information content (entropy) of
the system. Its ability to potentially take on different values is key. If
nothing changes, there is no time. We note that for one observer there might
potentially be more than one time-flow because there is more than one system
being measured. This might lead to the perception that, even if nothing
changes with respect to a specific system, time is flowing nevertheless
because the feeling relates to another flow. Likewise, if there is no energy
available (in form of temperature in our model) to process information there
is no thermal time either.

Finally, imagine a universe with a very high temperature and without any
interactions yet. The beginning of (thermal) time would then coincide with
the first measurement induced erasure of information and would be
accompanied by a dissipation of very big amounts of energy. There should
also be a measurable background radiation caused by all the continuing
measurements in the universe. Note that if the observer is not assumed to be
quasi-classical, then a very strange world emerges, where change is
registered in random time intervals and potentially instantaneous action is
possible, and where clocks could not march in step, not even with
themselves\bigskip . \newline
1. Barbour J. The End of Time.\textit{\ Phoenix,} \textbf{1999.}\newline
2. Rovelli C. Relational Quantum Mechanics.\textit{\ Int.Theor.Phys.}, 
\textbf{1996}, 35,1637-1678.\newline
3. Rovelli C. Incerto tempore incertisque locis. \textit{Found .Phys.},%
\textbf{\ 1998}, 28, 1031\newline
4. Lubkin E. Keeping the Entropy of measurement : Szilard revisited.\textit{%
\ Int. J. Theor. Phys.},\textbf{\ 1987}, 26, 523-535.\newline
5. Plenio M.B.; Vitelli V. The physics of forgetting: Landauer's erasure
principle and information theory.\textit{\ Contemp. Phys.,} \textbf{2001},
42, 25-60.\newline
6. Margolus N.; Levitin L. The maximum speed of dynamical evolution.\textit{%
\ Physica D,} \textbf{1998,} 120, 188-195.\newline
7. Cerf N.J.; Adami C. Negative entropy and information in quantum
mechanics. \textit{PRL 79},\textbf{1997}, 5194-99\newline
8. Rovelli C. Forget time.\textit{\ Essay for FXQI contest}, \textbf{2008}.%
\newline
9. M. Planck,\textit{\ Ann.Phys.} Leipzig,vol.6,\textbf{\ 1908}\newline
10. Mermin D. What is quantum mechanics trying to tell us?, \textit{Am. J.
Phys.},\textbf{98}, 66, 753-67

\end{document}